\documentclass[12pt]{article}

\usepackage{graphicx}
\newcommand{\be}{\begin{equation}}
\newcommand{\ee}{\end{equation}}
\newcommand{\bea}{\begin{align}}
\newcommand{\eea}{\end{align}}

\def\({\left(}
\def\){\right)}

\usepackage{color}
\usepackage{multirow}
\usepackage{graphicx}
\usepackage{hyperref}
\usepackage{amsmath}
\usepackage{amssymb}
\usepackage{amsfonts}   
\usepackage{latexsym}   
\usepackage{slashed}
\usepackage{yfonts} 
\usepackage{verbatim}
\usepackage{wrapfig}

\begin{document}

\centerline{\large{A comment on effective field theories of flux vacua}}
\bigskip
\bigskip
\centerline{Shamit Kachru$^1$ and Sandip P. Trivedi$^2$}
\bigskip
\bigskip
\centerline{$^1$ Stanford Institute for Theoretical Physics}
\centerline{Stanford, CA 94305 USA}
\medskip
\centerline{$^2$Department of Theoretical Physics}
\centerline{Tata Institute for Fundamental Research}
\centerline{Colaba, Mumbai 400005, India}
\bigskip
\bigskip
\begin{abstract}

We discuss some basic aspects of effective field theory applied to supergravity theories which arise in the low-energy limit
of string theory.  Our discussion is particularly relevant to the effective field theories
of no-scale supergravities that break supersymmetry, including  those that appear in constructing de Sitter solutions of
string theory.

\end{abstract}

\newpage
\tableofcontents

\section{Introduction}

Effective field theory is a central tool in quantum field theory and string theory, particularly useful in systems which cannot
be solved exactly, and where many extraneous degrees of freedom are irrelevant to some of the questions of central interest.  Front
and center in the list of such systems are compactifications of superstring theory: the behavior of the excited string modes,
and the massive Kaluza-Klein modes associated with the compact extra dimensions, is often unimportant if one's goal is to
understand basic facts about the low-energy physics, such as the vacuum structure.

\bigskip
The power of effective field theory has been one of the reasons progress in studies of string compactification has
been possible, despite the formidable complexity of writing down full solutions to the theory.  For instance, even for the most famous
and best understood vacua of the theory -- vacuum solutions with $N \geq 2$ supersymmetry arising from Calabi-Yau compactification of type II strings --
full gravity solutions lie far beyond reach.  Yau's theorem \cite{Yau}
guarantees a solution of the Einstein equations, but explicit metrics are not known.  Furthermore, the metric guaranteed by
the theorem solves the Einstein equations, but not the Einstein equations with $\alpha^\prime$ corrections.  The argument that solutions to the full series of corrected equations
exists -- order by order in $\alpha^\prime$ -- is obvious with hindsight in 
effective field theory (following from the absence of a superpotential for gauge singlets in $N=2$ supersymmetric field theories), but required careful reasoning in the 1980s \cite{Sen}.
The story in $N=1$ supersymmetric Calabi-Yau compactifications of the heterotic string is more subtle.  A superpotential can be shown not to obstruct the existence of
finite radius solutions to all orders in perturbation theory 
\cite{Witten}, through an argument combining holomorphy with axion shift-symmetries;
but this can fail at the level of non-perturbative effects 
\cite{DSWW}.

\bigskip
The results of these analyses are now well known to all string theorists.  We recall the history here only to make it clear how central the use of effective field theory -- in understanding equations too intractable to solve exactly -- 
was in developing our basic picture of even the most well known classes of $N=2$ and $N=1$ supersymmetric Minkowski vacua of
the theory.

\bigskip
The case of non-supersymmetric, non-vacuum solutions -- most prominently, those incorporating background fluxes or non-perturbative sources -- leads
to basic new questions.  
A question which has been brought to our attention (c.f. \cite{SS, DVR}) is: ``how am I to think of finding stable vacua in a system that has runaway directions
in perturbation theory, but potentially important non-perturbative physics?" As we will recall, this is a natural issue to consider in the context of no-scale supergravity theories,
which appear in various limits of string theory.
In this note, through very simple examples, we seek to illustrate how to think about this kind of issue.  We expect our results are obvious to many experts, but questions and comments from various sources
led us to believe that writing this note might nevertheless serve a useful purpose.

\section{The Polonyi model}

We start with a classic example of supersymmetry breaking in supersymmetric quantum field theory, the Polonyi model.  This is reviewed nicely in \cite{Luty}.

\subsection{Basic facts}

\bigskip
Imagine an $N=1$ supersymmetric low-energy theory with a single chiral superfield $S$.  The low-energy action is specified by a
choice of K\"ahler potential $K$ and superpotential $W$.  
We choose, to start with,
\be
K = S^\dagger S
\ee
(the canonical K\"ahler potential), and
\be
W = \mu^2 S~.
\ee
The theory is free; it formally appears to break supersymmetry
\be
|F_S|^2 \sim \vert \mu^2|^2~,
\ee
but it enjoys no Bose-Fermi splittings.  There is a moduli space
of degenerate vacua, one for each value of the lowest component of S.
(We will henceforth adopt the common abuse of notation of denoting the scalar
component of a chiral multiplet by the same symbol as the multiplet itself.)

\bigskip
In any generic UV completion, there can be additional fields which
induce additional interactions of S.  Integrating them out would
generally produce corrections to $K$; if the high-scale physics
occurs at scale $M$, then generically
\be
K = S^\dagger S + {c\over 4M^2}(S^\dagger S)^2 + \cdots
\ee
where $c$ is some $O(1)$ constant, and $\cdots$ denotes higher
order terms suppressed by appropriate powers of $M$.

\bigskip
The physics now depends on the sign of $c$ (since the K\"ahler
metric enters in the potential function when contracting the auxiliary fields in the chiral multiplet, usually denoted by $F_S$
and $\bar F_S$).  With conventions as in \cite{Luty}, the fate
of the vacuum manifold is:

\medskip
\noindent
$\bullet$ For $c>0$, there is no (calculable) vacuum.  The theory
runs off to large values of $\langle S \rangle$, where a new
effective field theory needs to be determined.

\medskip
\noindent
$\bullet$ For $c<0$, there is a vacuum at $S=0$.  It breaks
supersymmetry and has non-trivial splittings.  The dimensionful
parameter governing splittings is
\be
{|\mu^2|^2 \over M^2}~.
\ee

\subsection{Stabilizing the runaway}

Let us further discuss the physics for $c > 0$.  In the theory as it stands, there is no
vacuum within the regime of calculability of the theory.  However, we could consider slight
modifications of the theory.  In this case, they will be contrived.  Nevertheless, they will serve
to illustrate a point which arises again in natural ways later in the note.

\bigskip
One possibility is to couple $S$ to an $SU(N)$ gauge superfield via a gauge coupling function
\be
\label{sta}
\int d^2\theta \left({1 \over g^2} + f(S/M)\right) {\rm tr}(W_\alpha W^\alpha)~.
\ee
Let us consider the expansion of this theory around $S=0$; for simplicity, we 
set 
\be
f(0) = 0.
\ee
If we imagine the $SU(N)$ sector to be a pure gauge theory (with no charged chiral multiplets), then 
below the scale $\Lambda_{SU(N)}$ it confines, and produces a gaugino condensate.  The coupling above
results in a shift to the effective low-energy superpotential for $S$:
\be
W = \mu^2 S + f(S/M)\Lambda_{SU(N)}^3 + {1\over g^2}\Lambda_{SU(N)}^3~,
\ee
where $\Lambda_{SU(N)}$ is the dynamical scale of the gauge theory with coupling $g$ (as, recall, we are
expanding our effective theory near $S=0$).

\bigskip
The resulting vacuum equation for supersymmetric vacua is
\be
\mu^2 + {\Lambda_{SU(N)}^3 \over M} f^\prime(S/M) ~=~0.
\ee
This will have solutions where
\be
f' = -{\mu^2 M \over \Lambda_{SU(N)}^3~}.
\ee

\bigskip
Let us imagine $f = {1\over 2} a({S\over M})^2 + \cdots$
with $a$ some constant of $O(1)$.
This is natural if we absorb the constant into $g$, and imagine a ${\mathbb Z}_2$ symmetry acting on $S$ which is
broken only by the spurion $\mu$.  Then the equation for supersymmetric vacua becomes
\be
 a S + \cdots = -{\mu^2 M^2 \over \Lambda_{SU(N)}^3}~.
\ee
For a range of dynamical scales such that
\be
{\mu^2 M^2 \over \Lambda_{SU(N)}^3} \ll M \to M \ll {\Lambda_{SU(N)}^3\over \mu^2}~,
\ee
this gives a solution for $S$ in which the higher powers in the ellipses in $f(S/M)$ should be negligible.
$\Lambda_{SU(N)}$ should also be below our cutoff, yielding a range
\be
\label{rangea}
\Lambda_{SU(N)} \ll M \ll {\Lambda_{SU(N)}^3 \over \mu^2}
\ee
where we find, by a reliable self-consistent analysis, a supersymmetric solution near the origin in field space.
It is easy to see that now, a $c>0$ correction in the K\"ahler potential at most causes small
shifts to the properties of the leading solution -- and not runaway behavior.

\bigskip
We should also be sure that the mass of the $S$ field about the vacuum is small enough that it can be consistently incorporated 
in an effective field theory with higher scales $\Lambda_{SU(N)}$ and $M$. That is, if we call this mass $m$, we need
\be
\label{cond1}m \ll \Lambda_{SU(N)} \ll M~.
\ee
From the form of $f(S/M)$ above, we see that 
\be
\label{cond2}
m \sim {\Lambda_{SU(N)}^3 \over M^2}.
\ee
That this falls in the desired range for the effective field theory to be valid is then
already guaranteed from the constraints we found above.
Happily, then, all conditions are satisfied in the range of parameters given in eq.(\ref{rangea}).

\bigskip
Let us recap.  In the normal Polonyi model with a quartic correction to the K\"ahler potential generated by high
scale physics at a scale $M$, there are two possible fates -- a non-supersymmetric vacuum at the origin, or a runaway --
and which occurs depends on the sign of the leading correction to $K$.  
Here, we see that coupling the light field to another sector, even one with ${\bf non-perturbative}$ dynamics, can
stabilise the runaway, and can in fact yield a (self-consistent, reliable) supersymmetric vacuum state of the low-energy
field theory.  In this vacuum state, K\"ahler potential corrections that were previously the leading corrections determining
the vacuum structure, can be accounted for by small shifts of the vacuum.

\bigskip
The reader could be puzzled: when $\mu \neq 0$, there is no stable perturbative vacuum around $S=0$. Is it valid to compute
the non-perturbative physics of the $SU(N)$ sector absent a stable solution at the perturbative level in $g$?  The point of
Wilsonian effective field theory is that physics is determined by dynamics as a function of energy scale, whatever the couplings
involved.  In the regime where we work, the fast modes associated with the non-perturbative physics of $SU(N)$ condensation
must be integrated out in the supersymmetric effective field theory valid at  energy scales below the scale  where the $SU(N)$ confines.  This then
goes into a self-consistent determination of what physics (supersymmetry-preserving or otherwise) happens at lower energy
scales.

\section{Heterotic strings and no-scale structure}

Now, we move on to examples from effective supergravity theories which are closer to the heart of string theorists.  We consider the heterotic string compactified
on a Calabi-Yau threefold $X$, with a non-trivial gauge connection parametrizing a vector bundle $V$ with
\be
c_1(V) = 0, ~~c_2(V) = c_2(TX)
\ee
embedded into one of the
two $E_8$s, and the other left untouched.  
The low-energy theory then has, among other fields, a pure $E_8$ supersymmetric gauge theory with coupling controlled (at
tree level) by the dilaton multiplet, conventionally denoted by $S$.

\bigskip
The heterotic string also allows for the activation of a three-form $H$-flux,  which generates a superpotential for complex structure moduli of the form
\be
W_{\rm flux} = \int_X H \wedge \Omega~,
\ee
with $\Omega$ the holomorphic three-form of $X$.
This superpotential depends only on the complex structure moduli of $X$, and for generic choices of $H $, it
fixes the complex structure. Neglecting the Chern-Simons terms in $H$, we have $H=dB\in H^3(X, {\mathbb Z})$.

\bigskip
The light fields remaining in a theory with flux of this sort will still include the dilaton multiplet $S$, the
volume modulus of $X$ which we can call $T$, and the $E_8$ gauge multiplet.
The superpotential, at energies below the scale of stabilization of the complex moduli (which is $\alpha^\prime \over R^3$ in terms
of the radius of $X$), and also below the scale of gaugino condensation in the hidden $E_8$, is given by
\be
W = c + \Lambda_{E8}^3~.
\ee
$c$ is a constant coming from the flux superpotential.  This system was first described in \cite{DRSW}.

\bigskip
The dimensional reduction from 10d supergravity tells us more.  We know the leading order K\"ahler potential for $S,T$ is
\be
K = - {\rm log} (S + S^\dagger) - 3 {\rm log} (T + T^\dagger)~.
\ee
Because of this detailed form of the K\"ahler potential, we see that for any superpotential $W = W(S)$ which is independent of 
$T$, one will have a nice cancellation in the supergravity potential
\be
V = e^K \left( \sum_{ij}g^{i\bar j}D_i W \overline{D_j W} - 3 {|W|^2 \over M_{pl}^2}\right)~.
\ee
The term involving $F_T = D_T W = ({\partial_T + {K_{,T} \over M_{pl}} })W$ will cancel against the $-3 |W|^2$, leaving
a positive semi-definite potential
\be
V = e^{K} g^{S\bar S} |D_S W|^2~.
\ee
Vacua of such a potential always lie at $V=0$.  They are supersymmetric precisely if $W$ vanishes in the vacuum. If $W \neq 0$ in
the vacuum, then 
$D_T W \neq 0$, and the vacuum is non-supersymmetric.  This is the hallmark of ``no-scale" supergravities \cite{noscale}; of course the existence of a flat direction (here, $T$) and non-supersymmetric vacua at
vanishing energy does not survive corrections to the leading no-scale picture.

\bigskip
This potential can be minimized to obtain stable values for the dilaton under the conditions that $W_{\rm flux}$ evaluated
at the vacuum for complex moduli of $X$ -- i.e., $c$ -- is small enough.  Otherwise, the dilaton values obtained are at strong coupling.
This poses somewhat of a challenge in the heterotic string, because the 3-form flux $H$ is {\bf real}, and tuning to find flux
choices which yield small $H \wedge \Omega$ given the integrality constraint seems likely to be impossible.
Of course the $T$ modulus also remains unfixed, in the no-scale approximation.\footnote{Various routes to further analysis of this 
system (including Chern-Simons terms in $H$ properly) were described in \cite{Gukov}, but the results did not seem abundantly promising, due to absence of small parameters.}

\bigskip
At any rate, let us analyze the physics of this system at the stage where it was left in \cite{DRSW} -- namely, with the no-scale
K\"ahler potential and superpotential given above.  The perturbative (in the dilaton) physics involves a constant $W$, 
$W \simeq c~.$ With this superpotential, the dilaton $F$-term leads to a runaway vacuum rolling towards weak coupling.

\bigskip
The inclusion of the $E_8$ condensate restores a sufficiently non-trivial perfect square structure to the scalar potential to allow
for non-trivial vacua, at vanishing energy in the leading approximation, at fixed values of the dilaton.

\bigskip
Why is it valid to include the $E_8$ condensate, with its dilaton-dependent coupling, when the tree-level physics has a runaway behavior?
The answer is as in our \S2.  It is necessary, in a Wilsonian treatment, to include all effects in $W$ which occur above your low-energy
cutoff.  At scales beneath the (field-dependent) $\Lambda_{E8}$, one should analyze the supersymmetric theory including the condensate.
The runaway of S can potentially be cured, just as the runaway of S was cured in the Polonyi model, by incorporating non-perturbative
physics to stop its runaway.  In any such valid construction, one would have to check for self-consistency, of course.

\bigskip
In this example, as $T$ is unfixed, this does not yield a stable vacuum (trustworthy or not), if $\langle W \rangle\ne 0$ and supersymmetry is broken.  But that is not the point; the point
is to explain, in the language of low-energy effective field theory, why the analysis of \cite{DRSW} and its heterotic M-theory lift in
\cite{Horava}, is correct.  It is valid to include the non-perturbative physics of the condensate in the superpotential, despite the
fact that the perturbative superpotential would have led only to runaway behavior.

\section{Type IIB flux vacua}

We now turn to a discussion of compactifications of type IIB string theory, which (for reasons of 
relative tractability) have been a focus of research on cosmological
solutions of string theory.

\subsection{Basic facts}

The effective 4d supergravity for IIB flux compactifications on (orientifolds of) Calabi-Yau threefolds $X$
was described in \cite{GKP}, building on earlier work of \cite{GVW} and \cite{DRS}.  The superpotential takes the form
\be
W = W_{\rm flux} + W_{np}
\ee
where 
\be
W_{\rm flux} = \int_X G_3 \wedge \Omega~,
\ee
$G_3 = F_3 - \tau H_3$ combines the RR and NS three-form fluxes $F_3$ and $H_3$ with a factor of the axiodilaton $\tau$, and
$W_{np}$ comes from non-perturbative effects.  These can include either D-brane instantons \cite{Witten2} or effects arising
from strong dynamics, both of whose existence is model-dependent, determined by the dynamics of the full underlying
compactification.  The non-renormalization theorem protecting this form of $W$ to all orders in perturbation
theory was proven in \cite{Fernando}.

\bigskip
In models with a single K\"ahler parameter $\rho$, which will suffice for our discussion, one has
\be
K = -3 {\rm log}(-i(\rho - \bar\rho)) - {\rm log}(-i(\tau - \bar \tau)) - {\rm log}(-i\int_X \Omega \wedge \bar\Omega))
\ee
at tree-level, and one sees again a no-scale structure, now in the limit where one restricts to $W = W_{\rm flux}$.  
Then the $|D_{\rho}W|^2$ cancels the $3|W|^2$ in the supergravity potential, and one finds
\be
V = e^K \left(\sum_{ij} g^{i\bar j} D_i W \overline{D_j W}\right)~,
\ee
where $i$ runs over the complex moduli of X which are projected in by the orientifold action, and the dilaton.

\bigskip
In the approximation that $W$ is given purely by the flux superpotential, supersymmetric vacua -- those with $D_i W = D_\rho W = 0$ -- appear with vanishing energy.
Their stability is robust against corrections to $K$, which of course will occur even in perturbation theory.  The equation
$D_\rho W = 0$ implies $W=0$, so the supersymmetric vacua have vanishing superpotential (evaluated in vacuum).

\bigskip
Non-supersymmetric vacua are those which have $D_i W = 0$ but 
\be
\left(W_{\rm flux}\right)\vert_{\rm vacuum} = W_0 \neq 0~,
\ee
 and so $D_\rho W \neq 0$ in the  vacuum. 
Integrating out the complex structure moduli and the dilaton, the low-energy theory has 
\be
W = W_0~,
\ee
\be
K = -3 {\rm log}(-i(\rho-\bar\rho))~.
\ee
This is before incorporation of any non-vanishing effects in $W_{np}$.  As is characteristic of no-scale supergravities,
$\rho$ remains a flat direction at this level.

\bigskip
At this step the reader may be puzzled.  Small corrections to $K$ will surely exist, suppressed by powers of $1/\rho$, and 
so the non-supersymmetric no-scale vacua -- and the flatness of the potential for $\rho$ -- will be spoiled.  Is there any excuse to consider
$W_{np}$ at all, even if effects predicted by the microscopics of string theory would appear therein and perhaps lead to
stable vacua upon inclusion?

\bigskip
We are back precisely in the situation encountered previously in \S2 and \S3.  Wilsonian effective field theory tells one that
physics is  organized by energy scale, not just by coupling constant expansions.  If dynamics relevant to $W_{np}$ occurs
at an energy above the energy cutoff of the  Wilsonian effective theory, one must integrate it out and incorporate its 
effects into the potential for low-energy fields.

\bigskip
This logic explains why superpotentials of the form
\be
\label{supa}
W = W_0 + A e^{ia\rho}
\ee
-- involving both a tree-level piece and non-perturbative terms -- can be used, as long as their use is self-consistent,
in systems like the low-energy limits of (suitable) IIB flux vacua.\footnote{In general, one would expect a complicated sum of instanton effects, perhaps including multi-covers; the simple approximation above would be valid only in special circumstances.  Similarly,
use of the leading K\"ahler potential would only be valid in special circumstances.}

\bigskip
There is an important new feature compared to the discussion of the similar superpotential in \S3.  Because the flux $G_3$ in IIB flux vacua is complex, the constant $W_0$ can be tuned -- roughly speaking, the two integral forms $F_3$ and $H_3$ can ``cancel'' to produce a small value of the constant in $W$, but still stabilise the complex structure (and dilaton).  This fact was used as crucial feature in
\cite{KKLT} to allow for tuning to obtain vacua of this kind of superpotential at reasonably large volume, and was quantified
more precisely in the statistical treatment of \cite{DD}.  The statistical treatment clearly implies that small $W_0$ can be attained.
At the values of the volume attainable for small $W_0$, the known K\"ahler potential corrections are sufficiently suppressed
by inverse powers of $\rho$ that they simply cause small shifts to the vacuum.  In other words, small $W_0$ allows self-consistency of the approximations, as discussed in \cite{KKLT}.

\subsection{Energy scales}

Our discussion above is telegraphic. 
Let us more carefully examine the various energy scales which arise in the construction of \cite{KKLT} and see why they self-consistently justify the  effective field theory used above for analysing the stabilisation of the $\rho$ modulus. 

\begin{figure}[h]
\centering
\includegraphics[width=\textwidth]{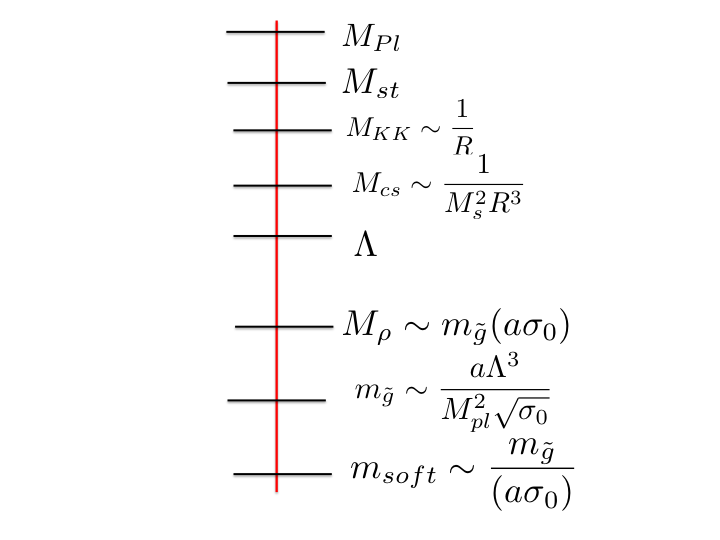}
\caption{Various energy scales in a class of IIB flux vacua. $M_{Pl}$: 4 dimensional Planck scale, $M_{st}$: string scale, $M_{KK}$: Kaluza-Klein mass, $R$: radius of compactification, $M_{cs}$: mass of complex structure moduli and dilaton, $\Lambda$: scale of non-perturbative dynamics, $M_\rho$: mass of $\rho$ modulus, $m_{\tilde g}$ mass of gravitino,  $m_{soft}$: soft susy breaking mass. $\sigma_0=R^4 M_{st}^4$, $a$ given in eq.(\ref{supa}).}
\end{figure}

\bigskip
Starting with the superpotential given in eq.(\ref{supa}), with small $W_0$, the various energy scales in the resulting supersymmetry preserving minimum with negative cosmological constant are given in Fig 1.  To simplify the discussion we have set $g_s\sim O(1)$ in Fig 1. (In addition we have  included the scale $m_{soft}$ which arises after supersymmetry breaking is included; see the discussion below.) A useful discussion of these scales appears in \cite{CN}.  Here we use the notation,
\be
\label{nca}
W_{np}=A e^{ia \rho} \equiv \Lambda^3,
\ee so that $\Lambda$ sets the scale of the non-perturbative dynamics. We also denote 
\be
\label{defisign}
\sigma_0= \langle Im (\rho) \rangle
\ee
where $\langle Im (\rho)\rangle$ is the value of the imaginary part of $\rho$ at the minimum. $Im (\rho)$  is related to the radius of compactification, $R$, by 
\be
\label{defr}
Im (\rho)= {R^4 M_s^4}.
\ee

\bigskip
It is easy to see from eq.(\ref{supa})  that the condition $D_\rho W=0$ leads to
\be
\label{sigmaa}
\sigma_0\simeq {1\over a} \ln\big( {-W_0 \over a A}\big).
\ee
Thus a small (negative) value of $W_0$ can give rise to a moderately large value of $\sigma_0$ and $R$ (in string units). 
This in turn  results in $\Lambda$ satisfying the condition
\be
\label{condl}
\Lambda \ll M_{Pl}.
\ee

\bigskip
If $m_\rho$ denotes the value of the $\rho $ modulus, the effective field theory used for analysing the stabilisation of  this  field  is valid  for
\be
\label{condkka}
M_{KK}, M_{cs},  \Lambda \gg m_\rho,
\ee
so that the KK modes,  complex structure moduli  and modes giving rise to the non-perturbative dynamics, are heavier degrees of freedom and can be integrated out. From Fig. 1 we see these conditions are self-consistently  met,  for moderately large $\sigma_0$,
 with 
 $\Lambda$ satisfying   eq.(\ref{condl}). 
 
\bigskip
The model discussed in \cite{KKLT} also incorporated a  supersymmetry-breaking sector, based on the study of
anti-D3 brane dynamics in a warped geometry in \cite{KPV}.
Including these effects gives rise to a potential\footnote{The scaling of the additional term is discussed around eqn (5.14) of \cite{KKLMMT}.}
\be
\label{potkklt}
V=e^K |D_\rho W|^2 + {D \over (Im \rho)^2}~.
\ee
Here $D$  is  determined by the warp factor of the throat where the anti-D3 brane is placed. Since the volume modulus has been stabilised before the breaking of supersymmetry was introduced, the addition of the anti-D3 brane leads to a metastable minimum for a range
of values of $D$. In fact, if $D$ roughly cancels the negative cosmological constant in the susy preserving $AdS$ vacuum referred to above, the shift in the vacuum expectation value of $ \rho$ is small, 
\be
\label{shdels}
{\delta \sigma_0 \over \sigma_0}\sim \big({1\over a \sigma_0}\big)^2 \ll 1 ~.
\ee
As a result the change in the values of $M_{KK}, M_{cs}, \Lambda$ and $m_\rho$  are all small  and eq.(\ref{condkka}) continues to be met. 

\bigskip
Once supersymmetry breaking occurs we need to make sure that one more condition is met \footnote{We thank Liam McAllister for urging us to be clear on this point.}. Namely, that  the soft masses which arise due to the supersymmetry breaking for  the degrees of freedom which are  being integrated out
are  small,  so that the resulting effective field theory is described by a supersymmetric Lagrangian to good approximation.  In Fig.1, $m_{soft}$  gives the soft mass which determines the mass splitting within  multiplets in the sector undergoing the non-perturbative dynamics. We see that it  is given by 
 \be
 \label{softm}
 m_{soft}\sim {\Lambda^3\over M_{pl}^2} {1\over \ a^{1/2} \sigma_0^{3/2}}.
 \ee
 As a result, as long as $\Lambda$ meets eq.(\ref{condl}) it also meets the condition
  \be
  \label{condf}
  \Lambda \gg m_{soft},
  \ee
  which ensures that the soft masses in this sector are small.

  \bigskip
  In this way we see that the use of the effective field theory that describes the stabilisation of the volume modulus is self-consistently justified, using as control parameters small $W_0$ and a low energy scale of supersymmetry breaking
(enabled here by use of a warped geometry).

 \subsection{Additional comments}

\bigskip
We have been focused in our discussion on one scenario, that of \cite{KKLT}, as it is a concrete model illustrating many
of the issues of interest in this note.
However, we should emphasize that other interesting classes of solutions arise
in the same general family of effective field theories, and are discussed in e.g. \cite{Large}.
These use  properties of corrections to the K\"ahler potential and are not reliant on tuning of the 
value of the flux superpotential.  \footnote{Far more general classes of constructions, differing in detail but similar in spirit in various ways
to those discussed above (most clearly in their application of self-consistent effective field theory), are described in \cite{Eva}.}

\bigskip
There has been rather extensive discussion of the properties of the set of such supergravities
that one can obtain from low-energy limits of string theory. Some non-perturbative aspects were studied in \cite{GKTT},  \cite{TT}, \cite{BK}.
Tests of aspects of the statistical treatment, by constructing large ensembles of vacua in the no-scale approximation, were performed in \cite{GKT, Conlon, DGKT}.  The statistical treatment captures the 
behavior of at least some aspects of the actual ensembles quite well.
Specific compactifications which seem to admit sufficiently rich instanton effects to admit
non-trivial solutions appear in \cite{DDF,DDFGK, Lust}.
A more thorough discussion of the considerable further work carried out in exploring the dynamics of this set of solutions can be found in the reviews \cite{MDSK,DDK,Denef}, as well as the textbooks \cite{McAllister,Uranga}.

\bigskip
It is also worth commenting on the connection of the discussion above with the ``no-go theorem" of Maldacena and Nunez \cite{MN} (as well as the earlier work of \cite{HD}). 
It was shown in \cite{MN}  that the equations of certain higher-dimensional {\it classical} supergravities do not allow for a de Sitter solution. 
We saw above that classical supergravity in the IIB theory after compactification on a Calabi-Yau orientifold gives rise to a no-scale theory with zero vacuum 
energy, which is in agreement with this result. However, what the effective field theory analysis also brings out clearly is that the volume modulus, $\rho$, is not 
stabilised due to the no-scale structure in the classical theory. This lack of stabilisation is in fact the chief obstacle for obtaining de Sitter vacua.  
Supersymmetry-breaking effects always scale as positive terms proportional to a power of $1/\rho$ in the scalar potential; this
will typically lead to runaway type behaviour, unless the string dynamics includes corrections to the classical supergravity which
stabilise $\rho$.  This tadpole for $\rho$ in the presence of positive energy sources is the main manifestation of the no-go theorems
in studying this class of vacua.

\bigskip
It follows then that to get around the Maldacena-Nunez argument, it is important to  find a way to stabilise the volume modulus. In the construction of \cite{KKLT}, the stabilisation of $\rho$ is  achieved by non-perturbative effects, as discussed above.  These go beyond classical supergravity (as do many other effects which are known to occur in string theory). Having achieved this stabilisation, the anti-D3 brane can introduce the breaking of supersymmetry, as in \cite{KKLT}, without destabilising the vacuum. 

\bigskip
Importantly, one can draw a broader lesson from this, independent of some of the details of scenarios like \cite{KKLT}.
This is that once the volume modulus is stabilised, supersymmetry breaking is not difficult to achieve. The scale of supersymmetry breaking does need to be kept low, compared to the string scale, since the volume modulus is being stabilised by non-perturbative effects. This is achieved  by placing the anti-D3 brane at the bottom of a warped throat in the very specific scenario of \cite{KKLT}, but more generally one expects to be able to achieve this in other ways as well.  Indeed, this is the defining characteristic of models of 
dynamical supersymmetry breaking.  Dynamical supersymmetry breaking is well known to happen in many examples; see e.g.  \cite{PT}, \cite{Shirman}, \cite{ISS}, \cite{Intriligator}. 

\bigskip
The comments above are not intended to indicate that our understanding of \cite{KKLT} and related constructions cannot be improved. 
There are several ingredients that need to come together in such constructions. One needs to stabilise the complex structure moduli and the dilaton with small $W_0$, eq.(\ref{supa});  have non-perturbative effects which stabilise the K\"ahler moduli; and finally  have a source of supersymmetry breaking. We have discussed above why these  are all important. Since the original discussion of these models, there has been considerable progress  leading to a better understanding of these effects.  We are now confident that they do arise.  To improve our understanding,  it will also be  worthwhile to have some explicit   examples showing that they can all be  simultaneously present in  string  compactifications. Work in this direction appeared in \cite{DDF,DDFGK,Lust}.

\bigskip
Finally, we note that  from a conceptual point of view, our understanding of metastable de Sitter vacua and the related picture of a landscape is still  at a very  preliminary stage. This is especially clear in comparison to our state of knowledge of their
AdS cousins.
Very clearly, more work can and should be done to improve this situation, and to make us fully confident of our understanding of supersymmetry breaking and de Sitter type space-times in string theory.

\section{Discussion}

In this note, we have addressed a question that has been raised about the treatment of (non-perturbative) quantum corrections in
the construction of superstring vacua.  We believe that the standard techniques of effective field theory, reviewed here, amply
justify the treatment in papers such as \cite{KKLT}, under the hypotheses specified there.  The addition of an anti-D3 brane
to create a metastable supersymmetry-breaking state \cite{KPV}, which is important in the effective field theory of 
\cite{KKLT}, has also been a subject of active discussion.  A thorough description
of how the effective theory treatment of the anti-brane is justified has already appeared 
in e.g. \cite{Joe}.\footnote{A useful analysis of some dramatic claims about the anti-brane effective field theory can also be found
in \cite{KLtoappear}.}

\bigskip
A recent paper \cite{OV}, motivated largely by no-go theorems with limited applicability to a {\it partial set} of {\it classical} ingredients, made a provocative conjecture implying that {\it quantum} gravity does not support de Sitter solutions.\footnote{Criticisms of this conjecture already appear in
\cite{Denefrecent}, \cite{Conlonrecent}.}  Our analysis
-- and more importantly, effective field theory applied to the full set of ingredients available in string theory --
is in stark conflict with this conjecture.  This leads us to  believe that the conjecture is false.

\bigskip
\centerline{\bf{Acknowledgements}}

\bigskip
We thank F. Denef, M. Douglas, A. Hebecker, R. Kallosh, A. Linde, L. McAllister, S. Shenker, E. Silverstein, and T. Wrase for valuable
discussions and comments.  The research of S.K. was supported in part by a Simons Investigator Award, and by the National Science
Foundation under grant NSF PHY-1720397.  S.T. is supported by a J.C. Bose Fellowship from the DST, the Infosys Endowment for Research on the Quantum Structure of Spacetime and by the DAE, Government of India.

\end{document}